\documentclass[pra,twocolumn,epsfig,rotate,showpacs]{revtex4}
\usepackage{amsfonts} 
\usepackage{amsmath}
\usepackage{amssymb}
\usepackage{bm}
\usepackage[usenames, dvipsnames]{color} 
\usepackage{dcolumn}
\usepackage{epic} 
\usepackage{epsfig}
\usepackage{graphicx}
\usepackage[abs]{overpic}
\usepackage[lofdepth,lotdepth,caption=false]{subfig}

\usepackage[breaklinks,colorlinks = true,linkcolor = blue,urlcolor=blue,citecolor=blue]{hyperref}
\usepackage{soul}
\usepackage{times}
\usepackage{ulem}
\usepackage{wrapfig} 
\usepackage{xy} 

\newcommand{\beq}{\begin{eqnarray}}
\newcommand{\eeq}{\end{eqnarray}}
\begin{document}
\title{Reinvigorating Excited State Coherences in $V$ Level Systems with the use of Trap States}
\author{Z.S. Sadeq}
\affiliation{Department of Physics, 60 St. George Street, University of Toronto, Toronto, M5S 1A7, Canada}
\date{\today}
\begin{abstract}
We investigate potential schemes to reinvigorate coherences between excited states in a $V$ level system under various excitation conditions. Coherent pulsed, cw and noisy pulsed excitation are considered and the existence of a trap state is shown to allow for the resurgence of excited state coherence in the two pulsed scenarios. With a short trap time (strong decoherence), it is possible to restart excited state coherence in the pulsed scenarios. In the case of cw excitation with relaxation back to the ground state, the coherences between the excited states do not vanish in the long time limit but instead approach a non-zero value. 
\end{abstract}
\email{sadeqz@physics.utoronto.ca}
\maketitle

\section{Introduction}
Intense interest in the emergence of quantum coherence in photosynthetic light harvesting complexes \cite{col,engel,mancal, flemingrev,seth} have initiated studies of the ways quantum coherence, established in a molecular system,  can be sustained. Quantum coherence is also intrinsic to several well known optical phenomena such as EIT and the laser \cite{orszag}. The key element required for the function of these cases is the coherence between excited states; for the cases mentioned, these states are not directly coupled to each other via the electric field. However, excited state coherence is often lost on a fast timescale in complex systems. 

In both atomic and biological systems, the importance of a trap or sink state, i.e. a state to which the excitation is always funneled, plays a huge role in the function and dynamics of these systems \cite{flemingrev}. For example, these traps in atomic systems can be states to which excited states decay via spontaneous emission. In the case of photosynthetic pigments, a trap state could represent a reaction center. 

We investigate the possibility to reinvigorate coherent excited dynamics by dissipating population from a photo excited state to another state through spontaneous emission or a non-radiative decay processes. This is motivated by biological processes such as the funneling of excitation to a reaction center during the excitation of a photosynthetic pigment protein complex. This could also represent processes in atomic or molecular physics, whereby excited states decay to lower energy states through relaxation processes or spontaneous emission. It is of interest to know if it is possible to devise a scheme that can rejuvenate the transient coherent response observed in short time regime \cite{zs1, zs2,timur2,timur2a}. In the context of biological light-matter interaction, for example, if excitation on one site has been dissipated or funneled to another excitation site, is it possible to produce a second transient coherent response by excitation of the initial site? That is, will the excitation be {}``felt again'' by the initial site? 

In its simplest form this is equivalent to excitation
from the ground state to several excited states, and these states
decaying into either the continuum or some other state not coupled
to the ground state by the field via some spontaneous emission or
non-radiative decay process. We choose to use the latter picture as
it provides a much more suitable numerical model.

In this paper, we investigate the coherences manifested between the excited states of a three level $V$ system under various excitation conditions and sink rates. We focus on various types of excitation, such as cw sources, coherent pulsed sources, as well as incoherent pulsed sources, in an effort to simulate both quantum optical experiments as well as biological light matter interaction \cite{flemingrev}. 

\begin{figure}[h]
\begin{center}
\includegraphics[scale=0.5]{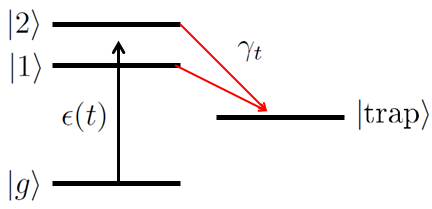}
\includegraphics[scale=0.5]{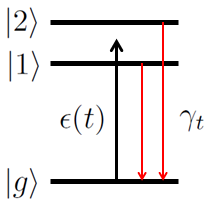}
\caption{(Color online) Two configurations investigated in this paper are as follows: a non degenerate $V$ type system, starting off in the ground state, is excited to the two excited states whose population is subsequently transferred to a lower energy trap state (top) and a non degenerate $V$ system but instead the population is transferred to the ground state (bottom). The bold black arrow represents the electric field while the red arrow represents the process that transfers population into the trap states. The rate at which this transfer occurs is given by $\gamma_{t}$.}
\label{schmt}
\end{center}
\end{figure}

We  investigate whether or not it is possible
to maintain coherences between excited states by transferring population
either to a trap state or to the ground state. To this end, we study
a non degenerate $V$ type atom which, starting off in the ground state,
is excited to the two excited states via pulsed, cw and noisy pulsed excitation. Population
is subsequently transferred into a lower energy trap state or to the ground state. These schemes are illustrated in Fig \ref{schmt}. 

This paper is organized as followed: Sec II covers the model used in this paper, Sec III explores coherent pulsed excitation, cw excitation is considered in Sec IV and noisy pulsed excitation is investigated in Sec V. The paper is then concluded in Sec VI. 
                                                                                              
\section{Model}
The system is taken to be a three level atom in the non degenerate $V$ configuration (see Fig \ref{schmt}). It is assumed that the original system Hamiltonian, $H_{0}$, is already diagonalized. The initial state is chosen as the atomic ground state, i.e. $\rho(0) = |g\rangle\langle g|$. Interaction with light is treated semi-classically via the electric dipole approximation. The field couples the two excited states, designated as $|1\rangle$ and $|2\rangle$. 

The trap state, designated $|\text{trap}\rangle$, is included in the dynamics in a optical Bloch equation type manner \cite{schloss,brumerbook}. The transfer of population from the excited states to the trap states is Markovian.

The Hamiltonian of the three level system, in the presence of the electric field $\epsilon(t)$, is given by

\begin{equation}
H=\left(\begin{array}{ccc}
\hbar \omega_{g} & -\mu \epsilon(t) & -\mu \epsilon(t)\\
-\mu \epsilon^{*}(t) & \hbar\omega_{1} & 0\\
-\mu \epsilon^{*}(t) & 0 & \hbar \omega_{2}
\end{array}\right).
\end{equation}
We define $\hbar\omega_{i}$ as being the energy of state $|i\rangle$. $\mu$ is the transition dipole moment between the ground state and the excited states. We assume these are the same for both transitions $|g\rangle \rightarrow |1\rangle$ and $|g\rangle \rightarrow |2\rangle$. We then solve the Liouville-von Neumann equation, 

\begin{equation}
\frac{d\rho}{dt}=\frac{i}{\hbar}\left[\rho,H\right]-\mathcal{L}\rho,
\label{liou}
\end{equation}
exactly numerically. The relaxation and decoherence is denoted by the non-unitary term $\mathcal{L}$. The equations of motion for the populations and coherences of the levels are
\begin{equation}
\frac{d\rho_{11}}{dt}=\frac{i}{\hbar}\left(\mu \epsilon^{*}(t)\rho_{1g}^{*}-\mu \epsilon(t)\rho_{1g}\right)-\frac{\gamma_{t}}{2}\rho_{11},
 \\
\end{equation}

\begin{equation}
\frac{d\rho_{22}}{dt}=\frac{i}{\hbar}\left(\mu \epsilon^{*}(t)\rho_{2g}^{*}-\mu \epsilon(t)\rho_{2g}\right)-\frac{\gamma_{t}}{2}\rho_{22},
\end{equation}

\begin{equation}
\frac{d\rho_{tt}}{dt}=\frac{\gamma_{t}}{2}\left(\rho_{22}+\rho_{11}\right),
\end{equation}

\begin{equation}
\frac{d\rho_{12}}{dt}=i\left(\omega_{21}\rho_{12}+\frac{\mu \epsilon^{*}(t)}{\hbar}\rho_{2g}^{*}-\frac{\mu \epsilon(t)}{\hbar}\rho_{1g}\right)-\gamma_{t}\rho_{12}.
\label{cohlio}
\end{equation}

The excited state populations are denoted by $\rho_{11}, \rho_{22}$ and the coherences between them are $\rho_{12}$. The trap state population is $\rho_{tt}$. We relate the sink rates to the familiar $T_{1,2}$ times as: $T_{1} = 2/\gamma_{t}$ and $T_{2} = 1/\gamma_{t}$ and $T_{2} = 2T_{1}$. We define the excited state splitting as: $\omega_{21} = (E_{2} - E_{1})/\hbar$, where $E_{i}$ is the energy of state $i$. 

The parameters that are varied are the electric field $\epsilon(t)$ and the sink rate $\gamma_{t}$. The electric field we choose to write as
\beq
\epsilon(t) = \epsilon_{0} \overline{\epsilon}(t).
\label{efieldlab}
\eeq
The quantity $\epsilon_{0}$ is the field strength and $\overline{\epsilon}(t)$ is the temporal profile of the electric field. We choose a Rabi frequency of $\mu \epsilon_{0} / \hbar = 10 \text{ THz}$ unless otherwise specified. 

\section{Excitation Using Coherent Pulses}

The first scenario we explore is coherent pulsed excitation. The pulses are short enough in time to induce both transitions $|g\rangle \rightarrow |1\rangle$ and $|g\rangle \rightarrow |2\rangle$. The pulse duration is also smaller than, or on the order of, the sink time scale, $T_{s} = 1/\gamma_{t}$. The $V$ level system is then subjected to a pulse train composed of two equally weighted pulses each with $\tau_{p} = 10 \text{ fs}$, leading to approximately a pulse duration at full width half maximum of $17$ fs. 

The temporal profile of the pulse train is given by
\begin{equation}
\epsilon(t)=\epsilon_{0}\left(e^{-\frac{(t-t_{1})^{2}}{\tau_{p}^{2}}}+e^{-\frac{(t-t_{2})^{2}}{\tau_{p}^{2}}}\right),
\end{equation}
where the pulse centers are $t_{1} = 250$ fs and $t_{2} = 750$ fs and the field strength parameter is $\epsilon_{0}$. 
\begin{figure}[h]
\begin{center}
\includegraphics[scale=0.32]{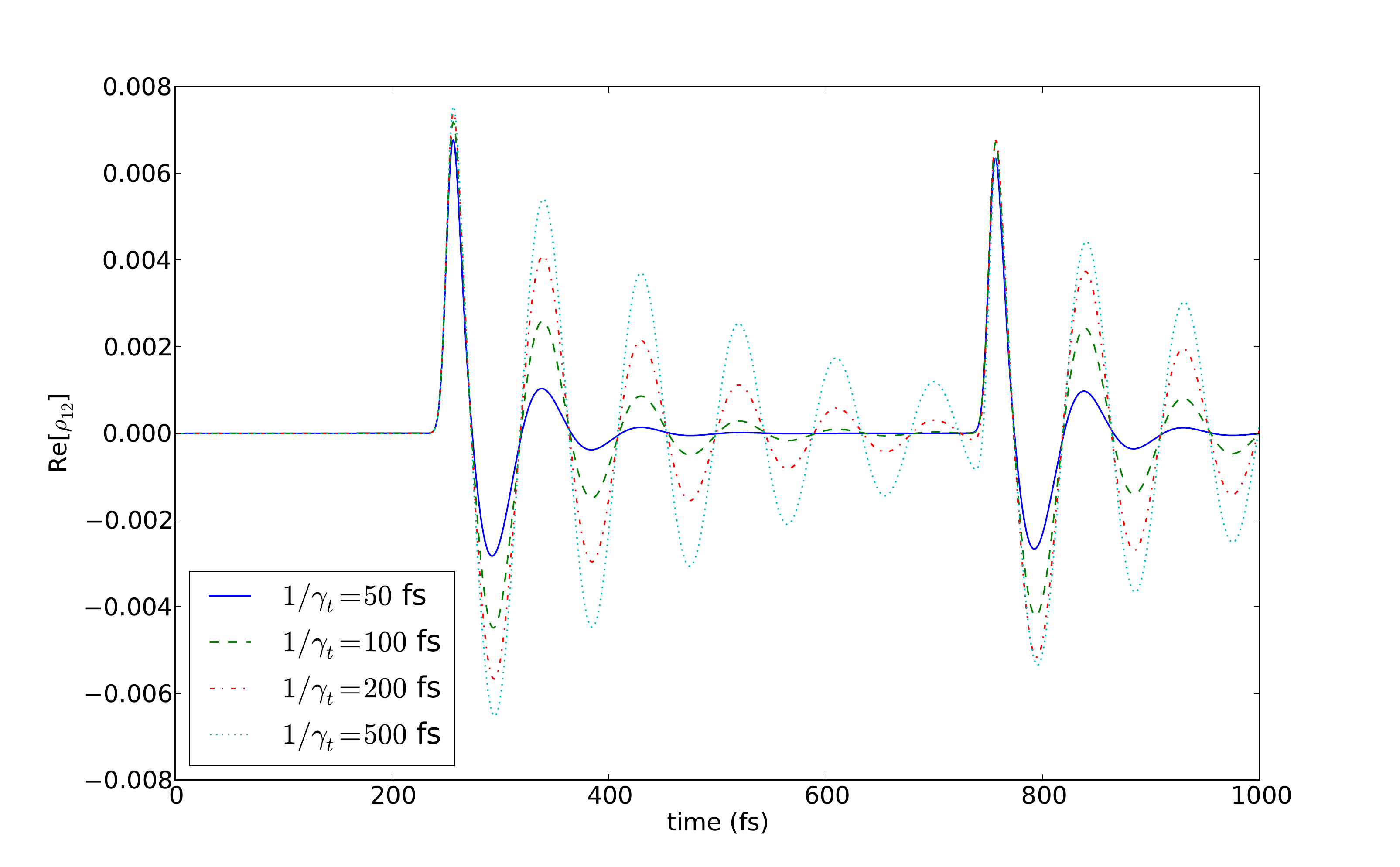}
\caption{(Color online) Coherences between excited states, $\rho_{12}$ for various sink rates $1/\gamma_{t}$. Pulse duration at full width half maximum is fixed at $17$ fs and the excited state period $\tau_{c} = 2\pi/\omega_{21}$ is set to $89$ fs.}
\label{cohpultrcoh}
\end{center}
\end{figure}

Both the excited state period and the pulse duration are fixed, but the sink time is varied. We solve the Liouville equation, Eq. \ref{liou}, for this system numerically. The results of this computation are presented in Fig \ref{cohpultrcoh}. 

From Fig \ref{cohpultrcoh}, it is clear that, for coherent pulses, it is possible to regenerate coherences between excited states for a wide range of rates of decoherence. However, for a weak rate of decoherence, i.e. $1/\gamma_{t}$ is large, the coherent response is much greater than that of strong decoherence (small $1/\gamma_{t}$). 

\begin{figure}[h]
\begin{center}
\includegraphics[scale=0.45]{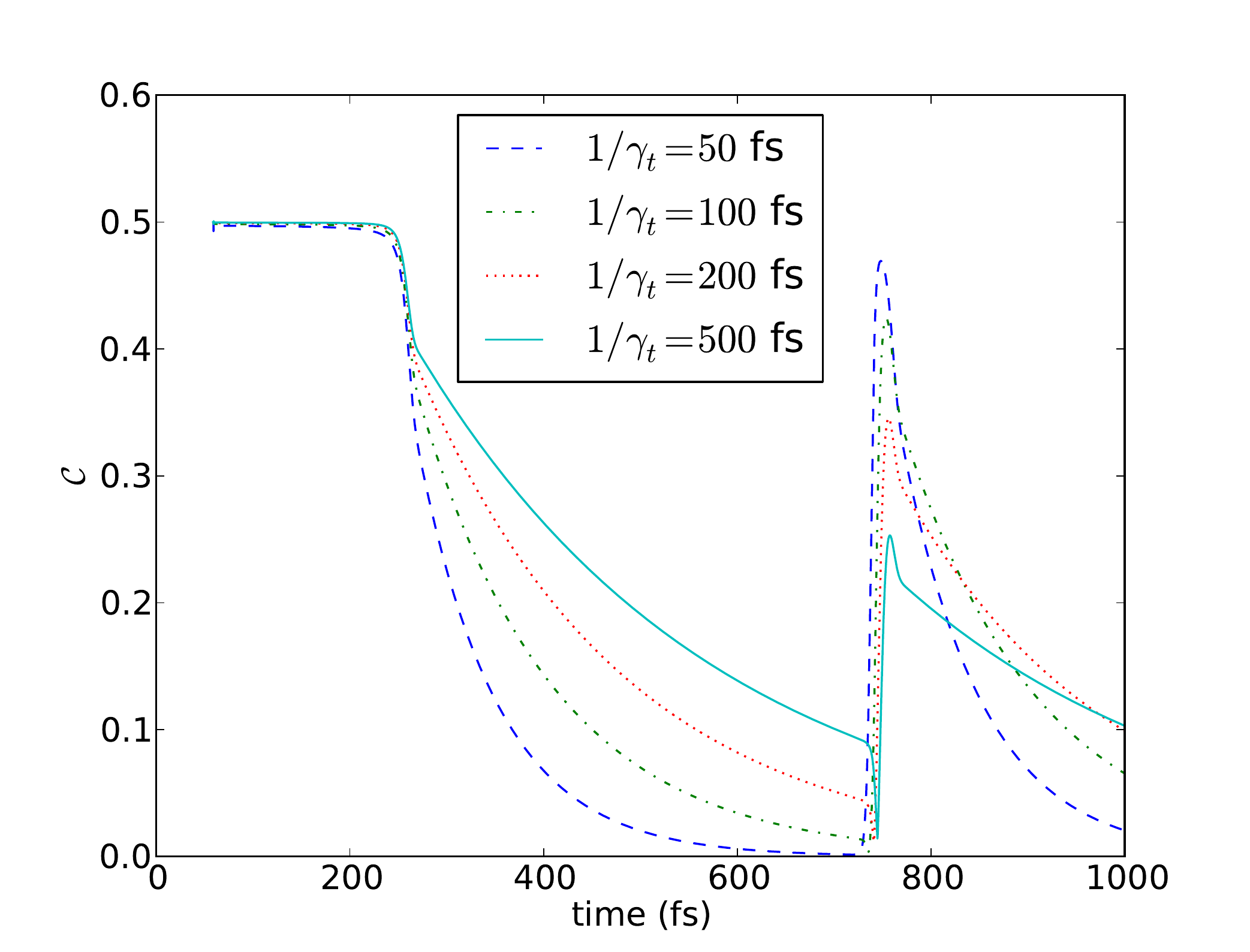}
\caption{(Color online) $\mathcal{C} = |\rho_{12}|/(\rho_{11}+\rho_{22})$, is plotted for various sink rates. Excited state period $\tau_{c} = 2\pi/\omega_{21}$ is chosen to be $89$ fs. $\mathcal{C}$ is used to elucidate excited state coherence fraction. For strong sink rates it was found that the resurgence of excited state coherence fraction is quite large upon the second excitation. Weak sink rates lead to a small resurgence of the excited state coherence fraction.}
\label{cohmthCtr}
\end{center}
\end{figure}

To quantify the strength of a coherent response, we use the measure $\mathcal{C} = |\rho_{12}|/(\rho_{11}+\rho_{22})$ from our previous work \cite{zs1,zs2}. This measure elucidates the mixed state nature of the excited state density matrix, which is a sub matrix composed of the excited state populations and the coherences between the excited states. This measure is \emph{independent} of the pump power, $\epsilon_{0}$. For the same system considered in Fig \ref{cohpultrcoh}, we calculate the value of this measure $\mathcal{C}$. These results are presented in Fig \ref{cohmthCtr}. 

It is clear from Fig \ref{cohmthCtr} that for very short sink times (large $\gamma_{t}$), the nature of the second coherent response will be a larger fraction of population than long sink times (small $\gamma_{t}$). Even though for weak decoherence, the second pulse generates more excited state coherence, it is a relatively small fraction of population than for the case of strong decoherence.

The existence of the trap states, which act to empty populations and coherences between the states, allows for a second excitation to rejuvenate coherences between excited states. Thus, in the coherent pulse excitation scheme it is possible to restart coherences. However, in the case of weak decoherence, the absolute value of excited state coherence is relatively high, it is still a small fraction of population. In order to achieve coherence with a large fraction of population, one requires a very large sink rate which quickly transfers the population into the trap state before the second pulse is applied. 

\section{cw Excitation with Relaxation to Ground State}
Consider, the case of cw excitation of a three level atom with a ground state, designated $|g\rangle$ and the two excited states $|1\rangle$ and $|2\rangle$. Here, the cw laser is in the center of the two excited state levels so as to excite states $|1\rangle$ and $|2\rangle$ transiently. In the short time regime, both states will be excited, their populations will be similar and the coherences between them, $\rho_{12}$, will be significant relative to their respective populations, $\rho_{11}$ and $\rho_{22}$. 

We consider the case whereby the excitation relaxes back down to the ground state, $|g\rangle$ instead of a trap state. We have investigated this scheme with the excitation being funneled to the trap state but we found that this scenario is unable to``restart" coherences between the excited state. This is explored in more detail in Appendix \ref{app:cwtrap}.

The difference between this case and the previous is the relaxation  transfers the excited state population to the ground state as opposed to a trap state. This modifies the equation for the ground state. The evolution of the ground state population is given by

\begin{widetext}
\begin{equation}
\frac{d\rho_{gg}}{dt}=\frac{i}{\hbar}\left(\rho_{1g}^{*}\mu \epsilon(t)+\rho_{2g}^{*}\mu \epsilon(t)-\rho_{1g}\mu \epsilon^{*}(t)-\rho_{2g}\mu \epsilon^{*}(t)\right)+\frac{\gamma_{t}}{2}\left(\rho_{22}+\rho_{11}\right).
\label{gtrapeq}
\end{equation} 
\end{widetext}
In the time evolution of the ground state population there are two processes happening, the pump from the ground
state to the excited states and the spontaneous emission from these excited states to the ground state. 

\begin{figure}[h]
\begin{center}
\includegraphics[scale=0.48]{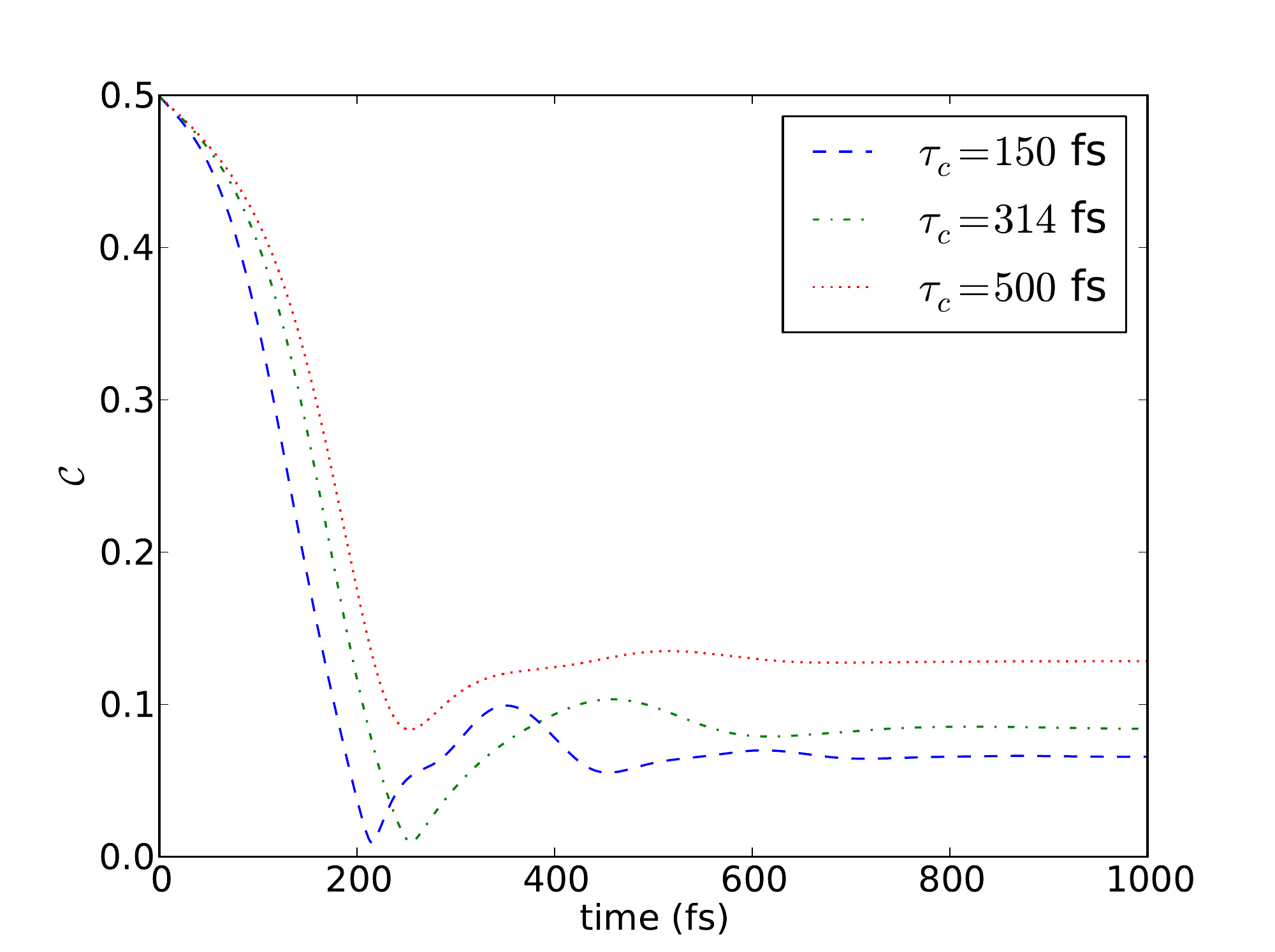}
\caption{(Color online) Mixed state measure, $\mathcal{C}$ plotted for various excited state splittings. As the splitting from the excited states get smaller, this value of steady state coherence becomes a larger fraction of population. Sink time is fixed for all three plots at $1/\gamma_{t} = 140$ fs. }
\label{gtrp}
\end{center}
\end{figure}

We solve the Liouville equation (Eq (\ref{liou})) for this model and present our findings in Fig \ref{gtrp}. In Fig \ref{gtrp} $\mathcal{C}$ is plotted for this system. The real and imaginary values of the excited state coherence, which are captured by the measure $\mathcal{C}$ can be found in Appendix \ref{app:cwgtrap}.

%
%

For smaller excited state splitting, i.e. in the limit of $\omega_{21}\rightarrow0$,
the coherences become a larger fraction of the excited state populations.
This is because the laser manages to excite both states $|1\rangle$
and $|2\rangle$, creating population and generating coherences between these two states. The coherence between the two states is sustained
by the equilibration between spontaneous emission and the pumping from the ground state to the excited states.
This allows the coherences between the excited states to be sustained in the sense that they are nonzero, but they do not evolve in time. 

\section{Excitation Using Noisy Pulses}

In this scenario we look at the coherences between the excited states, $\rho_{12}$, in a $V$ configuration with excitation using noisy pulses \cite{gardiner,brumerbook,brumer,zs1}. We utilize two noisy pulses whose statistics obey the following correlation function \cite{brumerbook}

\begin{equation}
\langle\epsilon(t')\epsilon^{*}(t'')\rangle=\epsilon_{0}^{2}e^{-\frac{\left(t'-t_{0}\right)^{2}}{\tau_{p}^{2}}}e^{-\frac{\left(t''-t_{0}\right)^{2}}{\tau_{p}^{2}}}e^{i\omega_{0}\left(t''-t'\right)}e^{-\frac{\left(t''-t'\right)^{2}}{2\tau_{d}^{2}}}.
\label{corr2}
\end{equation}

The details for the generation of noisy pulses are given elsewhere \cite{zs1}. A stochastic realization of the noisy pulse $\{ \epsilon(t) \}$ is generated, which is then subsequently used to generate a realization of the system response $\{ \rho(t) \}$. These responses are then collected and ensemble averaged to produce $\langle \rho(t) \rangle$.


\begin{figure}[h]
\begin{center}
\includegraphics[scale=0.4]{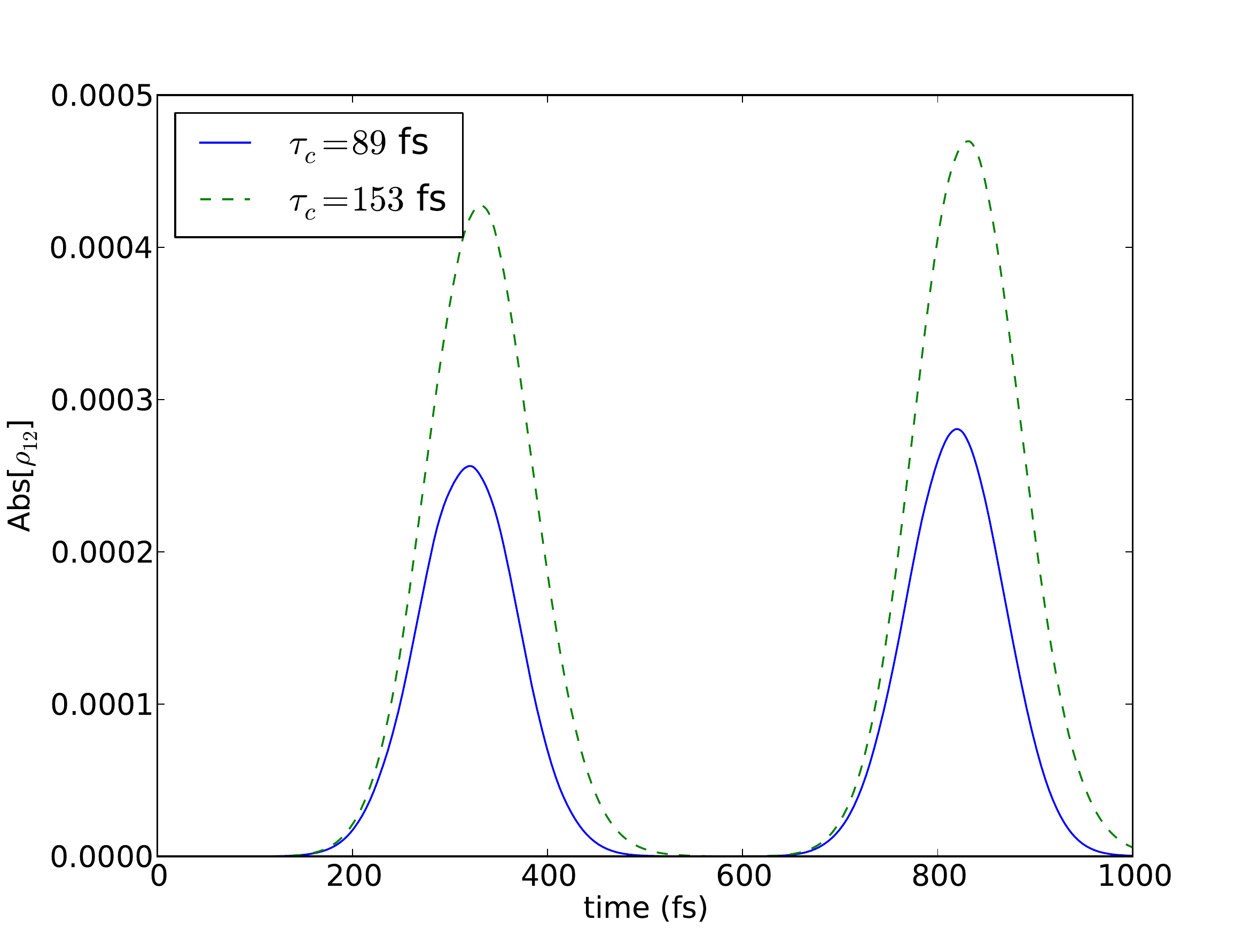}
\caption{(Color online) Comparison of coherences between excited states for two excited state splittings with a strong sink rate $1/\gamma_{t} = 20$ fs excited by two noisy pulses. First pulse has center at $t_{1}=50$ fs and second pulse has center at $t_{2}=550$ fs.   The Rabi frequency is: $\mu\epsilon_{0}/\hbar = 631$ GHz}
\label{noisytrap}
\end{center}
\end{figure}

\begin{figure}[h]
\begin{center}
\includegraphics[scale=0.4]{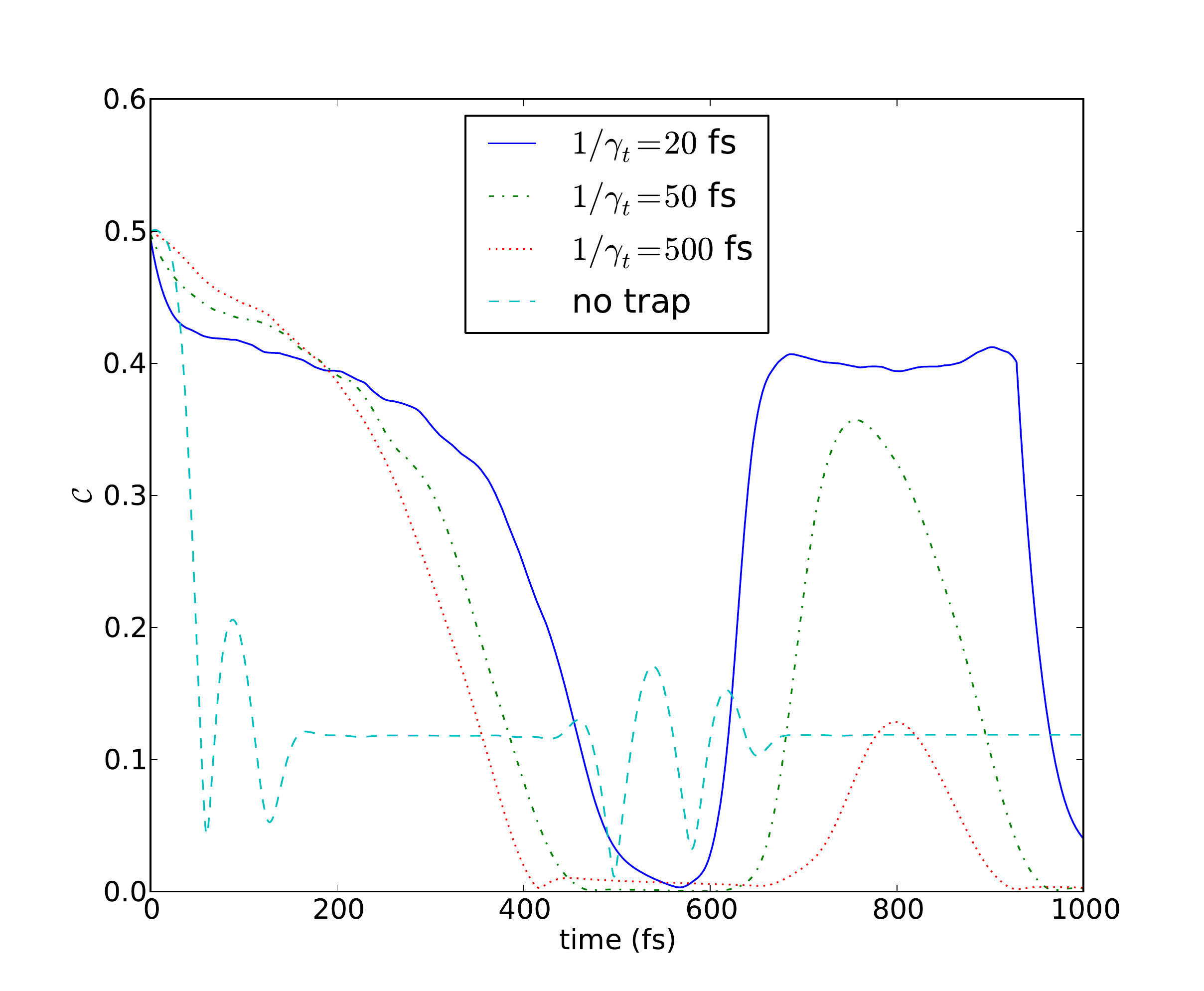}
\caption{(Color online) Excited state coherence as a fraction of population ($\mathcal{C}$) plotted for three level system with trap states (at various trap times) and no trap states for a fixed value of $\tau_{c}= 89$ fs. First pulse has center at $t_{0}=50$ fs and second pulse has center at $t_{0}=550$ fs.   The Rabi frequency is: $\mu\epsilon_{0}/\hbar = 631$ GHz}
\label{noisytrap2}
\end{center}
\end{figure}

The $V$ level system with the trap state is then subjected to a pulse train of noisy pulses. The pulse train is composed of two pulses centered at $t_{1} = 50$ fs and at $t_{2}=550$ fs. The Rabi frequency is set to $\mu\epsilon_{0}/\hbar = 631$ GHz. $\tau_{p} = 100$ fs so the full width at half maximum pulse duration was $167$ fs. The results of this calculation are presented in Fig \ref{noisytrap}.

There are now two decohering factors: 1) the population transfer process from the excited states into the trap state and 2) decoherence due to coupling to a noisy laser. By tabulating the coherences as a fraction of population as well as the absolute value of coherences, it can be determined precisely how coherent the response will be. 

It is evident from Figs \ref{noisytrap}, \ref{noisytrap2} that, through the influence of trap states and population draining, even under incoherent conditions it is possible to restart coherences between the excited states, $\rho_{12}$. If the excited state period becomes larger relative to the pulse duration, this magnitude of coherence becomes larger. It is apparent that these coherences exist only during the pulses and rapidly decay when the field no longer couples the ground to the excited states. The rejuvenation of coherence can be seen in Fig \ref{noisytrap} (top) where the second peak represents the second transient coherent response to irradiation with a noisy pulse. 


After the pulse is over, the coherent response quickly disappears due to the decoherence of the interaction with the pulse, as well as the decoherence associated with the population transfer to the trap state. This quick decoherence is determined by the ratio of excited state period $\tau_{c} = 2\pi/\omega_{21}$, the pulse duration $\tau_{p}$ and the sink time $1/\gamma_{t}$. 

For a $V$ level system with no trap, after each of the pulses the system experiences decoherence. This decoherence is manifested in the measure $\mathcal{C}$ in Fig \ref{noisytrap2}. However, unlike the case with a trap state, $\mathcal{C}$ does not approach zero. This is because even though the coherences are a small fraction of population, they are not funneled to another state. For systems with a trap state, the excitation is completely and irreversibly transferred to the trap state. This irreversible transfer of population and coherence to the trap state leads to $\mathcal{C} \rightarrow 0$. We must emphasize that this result is independent of the pump power $\epsilon_{0}$. While the magnitude of the coherences, as seen in Fig \ref{noisytrap}, is indeed proportional to pump power, the measure $\mathcal{C}$ is not. 

The presence of the trap, is essential for the resurgence of excited state coherence. The trap state ensures that population from excited states is transferred to the trap state. If this trap time is fast enough that all the population from the excited states is in the trap state then when the excitation occurs again, it will still be coherent. This can also be seen in Fig \ref{cohmthCtr} whereby strong trap times lead to a larger fraction of coherence. If the trap time is slow, then the second excitation occurs when the excited states are approaching a mixed state. Thus the added fraction of coherence of the second excitation is smaller. 

If there is no trap, the excited states are populated after the first pulse but the coherences between these states are small relative to their populations \cite{brumerbook, brumer}. After the pulse is over, the system approaches a mixed state. The second pulse generates populations and coherences, but the excited state coherence fraction is altered only slightly. For very weak coupling to the trap, the second response is sometimes less than the case with no trap states. Thus, it is important to not only have a trap state, but very strong coupling to the trap.

\section{Conclusion}
In this paper, we have examined scenarios where coherences between excited states can be reinvigorated with the use of trap states. This resurgence is possible in the pulsed case, where a large rate of decoherence leads to a stronger coherent response to the second pulse. In the case of cw excitation with population transfer to the ground state, the coherences approach a static non-zero value in the steady state. 

For the noisy pulsed case, it is possible to create a transient coherent response which exists for the time scale of the pulse. This coherent response decays quickly once the pulse is over, however, it can be restarted if the excited states are connected to a trap state. For a very strong coupling to the trap states it is possible to regenerate coherences. However, weak coupling to trap states does not yield a significant regeneration of coherences.

\begin{acknowledgments}
ZS would like to thank Y. Khan, T. Scholak, T. V. Tscherbul, V. V. Albert and R. Dinshaw for fruitful discussions.
\end{acknowledgments}

\newpage
\begin{widetext}
\appendix

\section{\label{app:cwtrap} $V$ System with Spontaneous Emission to Trap State}

\begin{figure}[h]
\begin{center}
\includegraphics[scale=0.4]{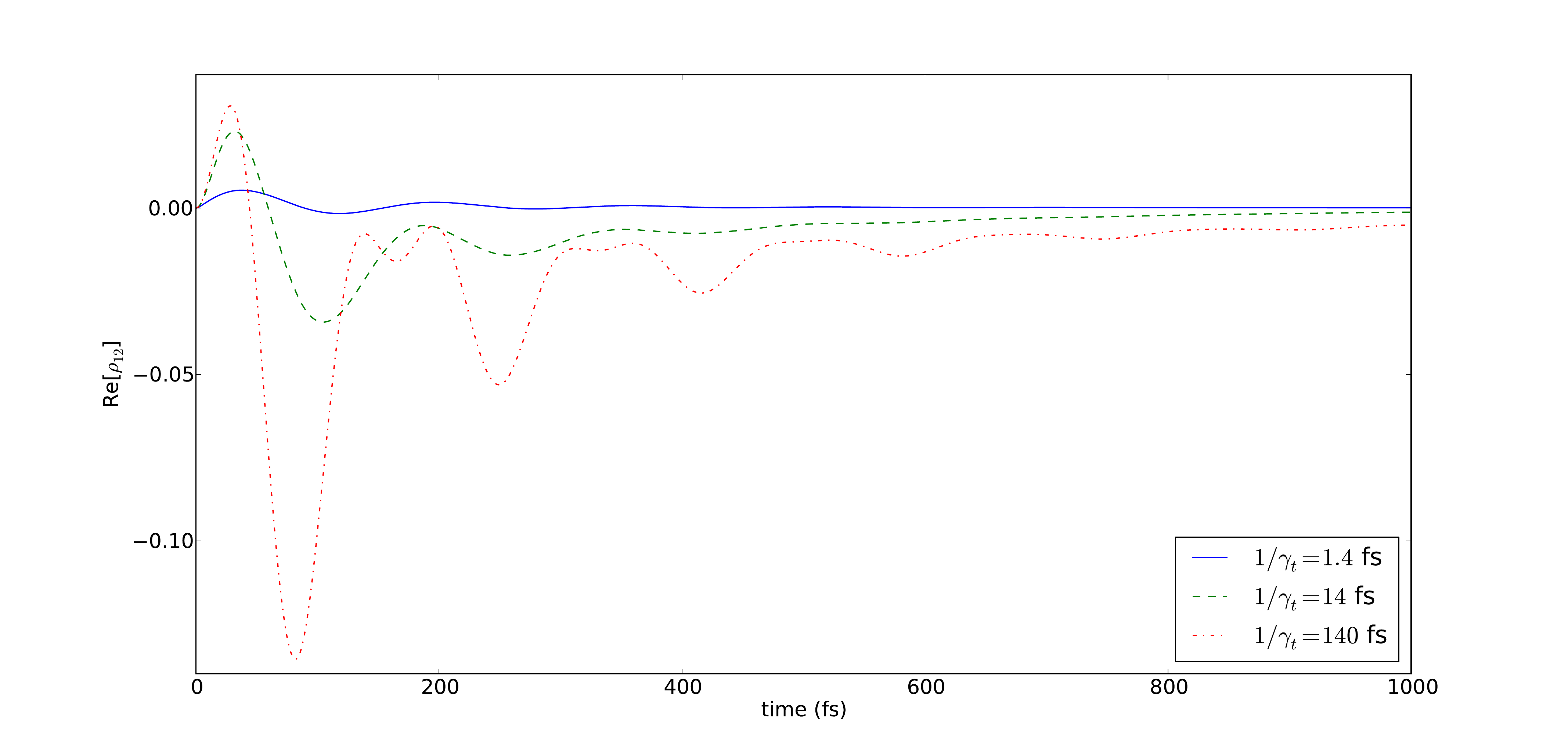}
\includegraphics[scale=0.4]{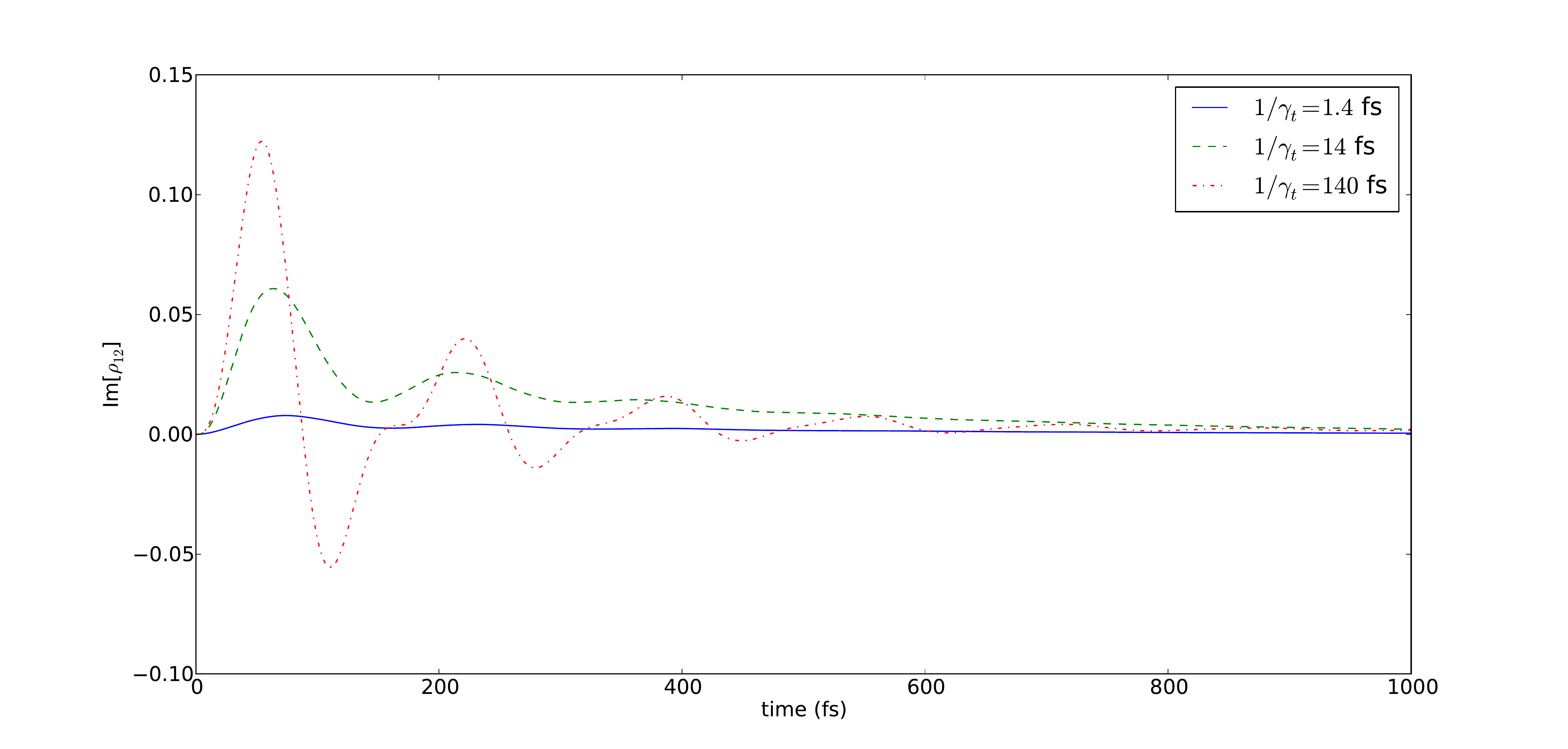}
\caption{(Color online) Re$\left(\rho_{12}\right)$ (top)  for various sink times. Im $\left(\rho_{12}\right)$ (bottom) for various sink times. $\tau_{c}=89$ fs. The system is irradiated with a cw source tuned to a frequency in between the two excited states $|1\rangle$ and $|2\rangle$. The excitation is then funneled to a trap state $|\text{trap}\rangle$. The Rabi frequency is $\mu\epsilon_{0}/\hbar = 10$ THz}
\label{cwtrp}
\end{center}
\end{figure}

In this section we examine a $V$ level system with a trap state $|\text{trap}\rangle$ like Fig. \ref{schmt} (top). The electric field used is a cw laser tuned in between the two states $|1\rangle$ and $|2\rangle$. We solve the Liouville-von Neumann equation for this system (Eq. \ref{liou}) numerically and tabulate the coherences between the excited states $\rho_{12}$ as a function of time. These are plotted in Fig. \ref{cwtrp}. 

From Fig \ref{cwtrp},  as population starts being transferred into the trap state, the dynamics of
the coherences between the excited states eventually decays to zero. Even though the laser can excite and populate both excited states transiently, the transfer of population to a trap state does not renew coherences and the system does not see the cw field
anew, unlike the pulsed case.

\section{\label{app:cwgtrap} $V$ System with Spontaneous Emission to Ground State}
In Section IV of the main text, cw excitation of a $V$ level system with relaxation to the ground state was explored. It was found that while the coherences are non zero in the long time limit, they do not evolve in time. We plot the real and imaginary values of the coherence in Fig. \ref{gtrp2} below. 
\begin{figure}[h]
\begin{center}
\includegraphics[scale=0.45]{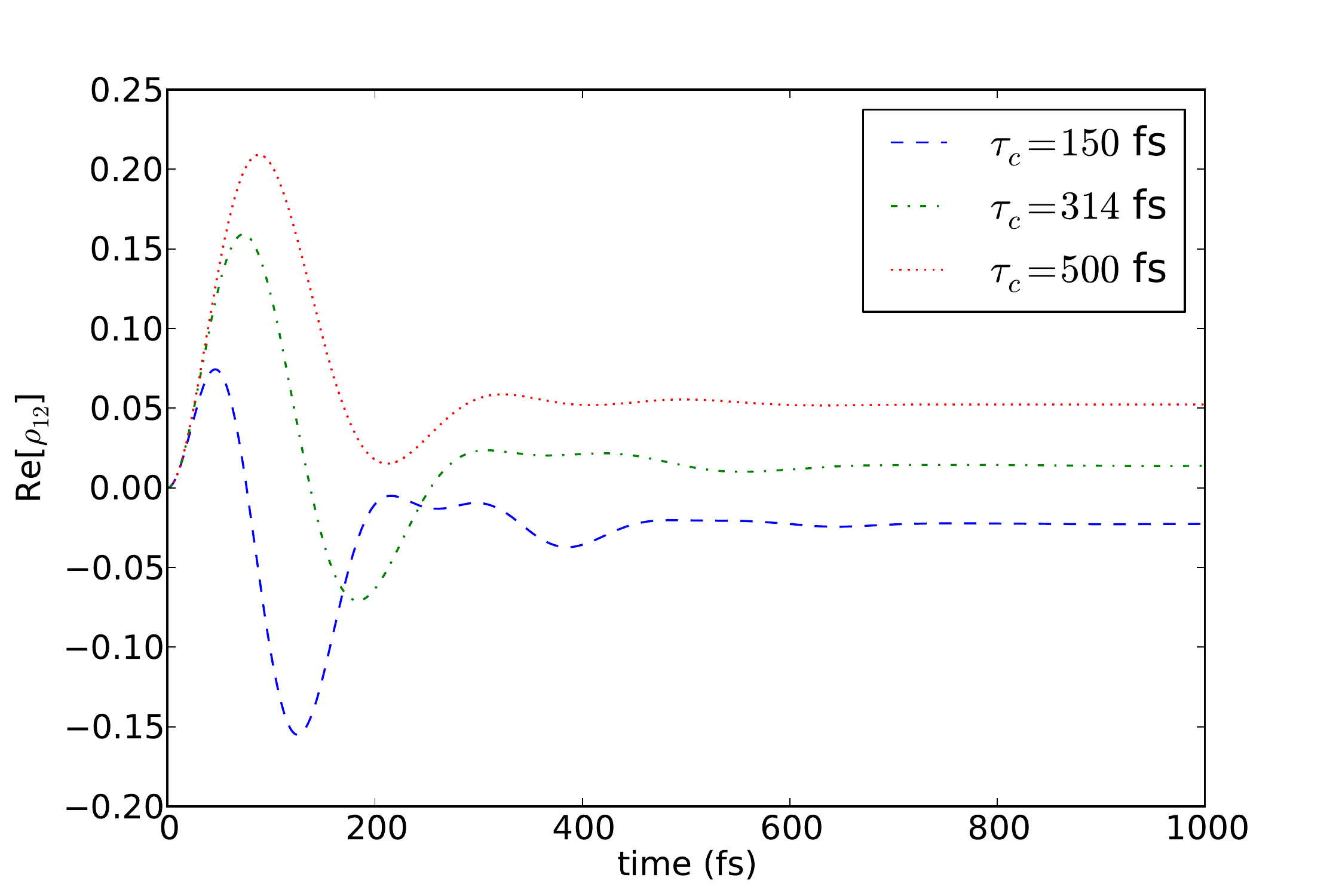}
\includegraphics[scale=0.45]{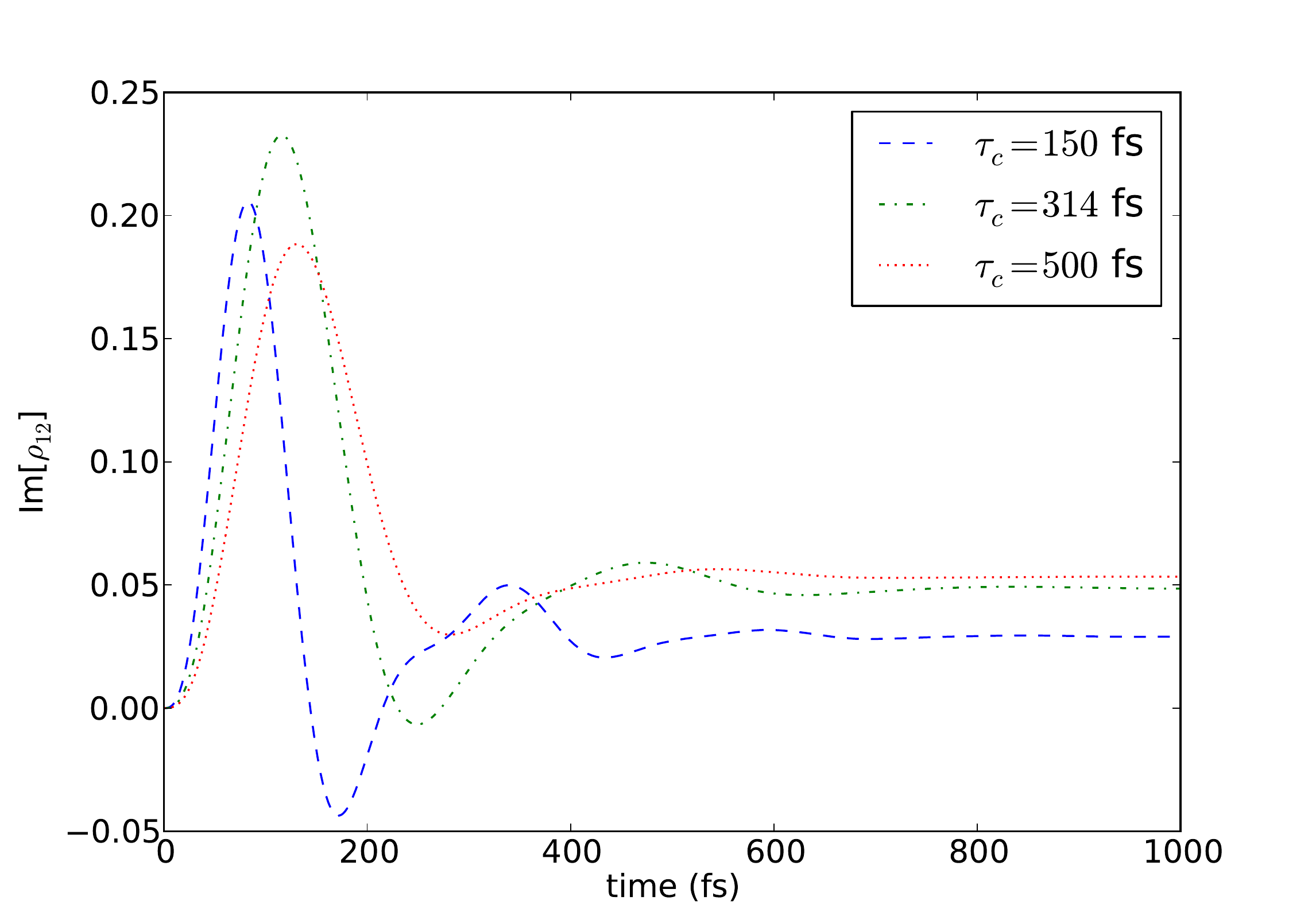}
\caption{(Color online) Re[$\rho_{12}$] (top) and Im[$\rho_{12}$] (bottom) whereby the excitation from the excited states is transferred into the ground state. Note in the steady state, both the real and imaginary components reach a non zero value. As the splitting from the excited states get smaller, this value of steady state coherence becomes larger. Sink time is fixed for all three plots at $1/\gamma_{t} = 140$ fs. The Rabi frequency is $\mu\epsilon_{0}/\hbar = 10$ THz.}
\label{gtrp2}
\end{center}
\end{figure}
\end{widetext}

\end{document}